\documentclass[twocolumn,pre,superscriptaddress,notitlepage]{revtex4-1}

\usepackage[pdftex]{graphicx}
\usepackage{amsmath}
\usepackage{bbm}
\usepackage{amssymb}
\newtheorem{definition}{Definition}
\usepackage{color}
\usepackage{enumitem}

\newtheorem{theorem}{Proposition}

\begin{document}

\title{Graph-based semi-supervised learning for relational networks}
\author{Leto Peel}
\email[]{leto.peel@uclouvain.be}
\affiliation{ICTEAM, Universit\'{e} catholique de Louvain, Louvain-la-Neuve, Belgium}
\affiliation{naXys, Universit\'{e} de Namur, Namur, Belgium}

\begin{abstract}
We address the problem of semi-supervised learning in relational networks, networks in which nodes are entities and links are the relationships or interactions between them. Typically this problem is confounded with the problem of graph-based semi-supervised learning (GSSL), because both problems represent the data as a graph and predict the missing class labels of nodes. However, not all graphs are created equally. In GSSL a graph is constructed, often from independent data, based on similarity. As such, edges tend to connect instances with the same class label. Relational networks, however, can be more heterogeneous and edges do not always indicate similarity. For instance, instead of links being more likely to connect nodes with the same class label, they may occur more frequently between nodes with different class labels (\textit{link-heterogeneity}). Or nodes with the same class label do not necessarily have the same type of connectivity across the whole network (\textit{class-heterogeneity}), e.g. in a network of sexual interactions we may observe links between opposite genders in some parts of the graph and links between the same genders in others. Performing classification in networks with different types of heterogeneity is a hard problem that is made harder still by the fact we do not know a-priori the type or level of heterogeneity. In this work we present two scalable approaches for graph-based semi-supervised learning for the more general case of relational networks. We demonstrate these approaches on synthetic and real-world networks that display different link patterns within and between classes. Compared to state-of-the-art baseline approaches, ours give better classification performance and do so without prior knowledge of how classes interact. In particular, our \textit{two-step label propagation} algorithm gives consistently good accuracy and precision, while also being highly efficient and can perform classification in networks of over 1.6 million nodes and 30 million edges in around 12 seconds.
\end{abstract}
  
  \maketitle

\section{Introduction}
Semi-supervised learning is a classification problem that aims to make use of unlabelled data, as well as the labelled data typically used to train supervised models.  A common approach is graph-based semi-supervised learning (GSSL)~\cite{belkin2004semi, joachims2003transductive, talukdar2009new, zhou2004learning, zhu2003semi}, in which (often independent) data are represented as a \textit{similarity graph}, such that a vertex is a data instance and an edge indicates similarity between two instances.  By utilising the graph structure, of labelled and unlabelled data, it is possible to accurately classify the unlabelled vertices using a relatively small set of labelled instances.
%
%

Here we consider semi-supervised learning in the context of \textit{relational networks}.  These networks are a type of graph that consist of nodes representing entities (e.g.~people, user accounts, documents) and links representing pairwise dependencies or relationships (e.g.~friendships, contacts, references). Here class labels are discrete-valued attributes (e.g.~gender, location, topic) that describe the nodes and our task is to predict these labels based on the network structure and a subset of nodes already labelled.  
  This problem of classifying nodes in networks is often treated as a GSSL problem because the objective, to predict missing node labels, and the input, a graph, are the same.  Sometimes this approach works well due to assortative mixing, or homophily, a feature frequently observed in networks, particularly in social networks.  Homophily is the effect that linked nodes share similar properties or attributes and occurs either through a process of selection or influence.  However, not all node attributes in relational networks are assortative.  For example, in a network of sexual interactions between people it is likely that some attributes will be common across links, e.g.~similar demographic information or shared interests, but other attributes will be different, e.g.~links between people of different genders.  Furthermore, the pattern of similarity or dissimilarity of attributes across links may not be consistent across the whole network, e.g.~in some parts of the network links will occur between people of the same gender.

\begin{figure}
  \includegraphics[width=\columnwidth]{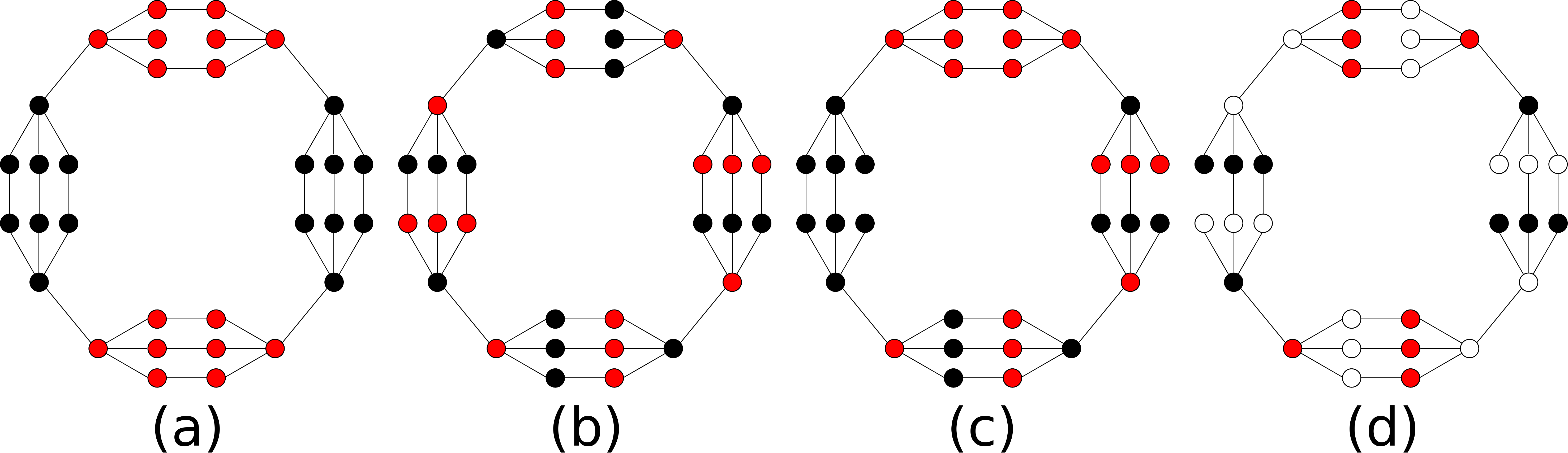}
  \caption{Different patterns of links between class labels \{red, black\}: (a) nodes with the same label tend to be linked (\textit{assortative}), (b) links connect nodes with different labels (\textit{link-heterogeneity}), (c) some nodes are assortative and some are not (\textit{class-heterogeneity}), (d) missing labels (white) obscures the pattern of links.}
  \label{fig_links}
\end{figure}

In situations where we have a sparsely labelled network and
do not know the pattern of interaction between nodes of different classes, the problem of predicting the class labels of the remaining nodes is hard.  Figure~\ref{fig_links} shows a toy example in which nodes are assigned red or black labels and Fig.~\ref{fig_links}(a)--(c) show possible arrangements of labels that become indistinguishable if certain labels are missing (Fig.~\ref{fig_links}(d)).  
Tasks such as fraud detection face this type of problem, where certain patterns of interaction are indicative of nefarious behaviour (e.g.~in communication~\cite{cortes2002communities} or online auction~\cite{chau2006detecting} networks) but only a sparse set of confirmed fraudulent or legitimate users are available and no knowledge of how fraudsters operate or if there are different types of fraudulent behaviour.    

In this work, we describe the problem of semi-supervised learning in relational networks (Sec.~\ref{sec_problem_def}) and present two novel methods for solving it.  Both methods approximate equivalence relations from social network theory to define a notion of similarity that is robust to different patterns of interaction (Sec.~\ref{sec_similarity_in_networks}). We use these measures of similarity to construct similarity graphs from relational networks upon which we can propagate class label information (Sec.~\ref{sec_methods}).  We demonstrate on synthetic networks that our methods are capable of classifying nodes under a range of different interaction patterns in which standard GSSL methods fail (Sec.~\ref{sec_results}).  Finally, we demonstrate on real data that our \textit{two-step label propagation} approach performs consistently well against baseline approaches and easily scales to networks with $>10^6$ nodes and $>10^7$ edges.

\section{Problem definition}
\label{sec_problem_def}
%
A relational network is represented as a graph $\mathcal{G}_r = \{\mathcal{V}_r,\mathcal{E}_r\}$ comprised of a set of $n$ vertices $\mathcal{V}_r$ and a set of $m$ edges $\mathcal{E}_r$.  The edges of the graph are unweighted and can be either directed or undirected.  Each node $i \in \{1, ..., n \}$ is assigned a class label $y_i \in \{1,...,\ell\}$, but we only observe these for a subset of nodes.
%
\begin{definition}{Network node classification} ---\\
\label{def_nodeclass}
Given a set of labelled nodes, $\mathcal{L}$, and the structure of the graph $\mathcal{G}_r$, find the class label $y_i$ for the remaining nodes $i \in \mathcal{U} : \mathcal{U} = \mathcal{V} \setminus \mathcal{L}$.
\end{definition}
This problem closely resembles that of graph-based semi-supervised learning (GSSL). 
Both have the same type of input (a graph) and output (labels on nodes). The crucial difference is what the input graphs represent, a seemingly minor detail that is often overlooked and the two problems treated the same~\cite{bertoni2011cosnet, bilgic2010active,gallagher2008using,lin2010semi, shin2006prediction,you2010semi}.  
However, the type of graph is important with respect to the assumptions we make in constructing a classifier.  GSSL methods are designed to be applied to a similarity graph $\mathcal{G}_s = \{\mathcal{V}_s,\mathcal{E}_s\}$, usually constructed from independent and identically distributed (iid) data, in which edges represent pairwise similarity between the nodes.  Therefore, in GSSL it is reasonable to make a smoothness assumption that connected nodes are more likely to be of the same class (assortativity). 
However, in relational networks, where the links represent interactions or relationships between the nodes, the distribution of class labels over the network may be more complex than simple assortativity. In the remainder of this section we describe the 
GSSL problem in more detail and describe types of heterogeneity in relational networks that can cause problems for GSSL approaches. 

\subsection{Graph-based semi-supervised learning}
\label{sec_gssl}
A common assumption in supervised learning is that similar instances are assigned the same class label, i.e.~class labels vary smoothly over instances embedded in a feature space.
This smoothness assumption means we can predict unlabelled instances based on the closest labelled instances, e.g. the k-nearest neighbours classifier.  However, when few labelled instances are available, supervised methods can perform poorly 
particularly when measurements are noisy. Semi-supervised learning addresses this issue by making a more global smoothness, or ``cluster'', assumption that data clustered together belong to the same class. GSSL captures the structure of the data by 
linking similar instances to form a graph. 
Thus, groups of nodes that are more densely connected are more likely to have the same class label.  

GSSL assigns a label score $F_{ic}$ for each node $i$ belonging to a class $c$, such that the label scores vary smoothly over the graph structure and remain consistent with any known label information.  Mathematically, this as an optimisation of the following form:
\begin{equation}
 \min_{\mathbf{F}} \quad \frac{1}{2} \rm{tr}(\mathbf{F^TLF}) + \lambda \| \mathbf{F-B} \|^2 \quad
 \textrm{s.t.} \quad \mathbf{F}\mathbbm{1} = \mathbbm{1} \enspace, \label{eq_gssl_opt} 
\end{equation}
%
where $\mathbf{L}$ is a linear operator on the adjacency matrix $\mathbf{A}$ of the graph, $\mathbf{F}$ is an $n \times \ell$ matrix of class labels scores, $\mathbf{B}$ is an $n \times \ell$ matrix that encodes our prior knowledge of class labels ($B_{ic}$ equals $1/\ell$ if node $i \in \mathcal{U}$, $1$ if the node $i \in \mathcal{L}$ and in class $c$, or 0 otherwise). 
The first term in Eq.~\eqref{eq_gssl_opt} ensures that label scores vary smoothly over the graph, while the second term 
is a fitting constraint to maintain consistency between label scores, $\mathbf{F}$, and initial label assignments $\mathbf{B}$.   The parameter $\lambda$ is a regularisation constant that allows us to adjust the balance between these two constraints.  

Different choices of $\mathbf{L}$ and $\lambda$ allow us to recover specific GSSL methods, for example setting $\lambda=0$ and using the graph laplacian $\mathbf{L} = \mathbf{D-A}$, where $\mathbf{D}$ is a diagonal matrix of node degrees ($D_{ii} = \sum_j A_{ij}$) recovers the harmonic function method~\cite{zhu2003semi}. Most relevant to this work is the setting where $\mathbf{L} = \mathbf{D}^{-\frac{1}{2}}\mathbf{AD}^{-\frac{1}{2}}$ and $\lambda=(\frac{1}{\alpha}-1)$.  Substituting these values into Eq.~\eqref{eq_gssl_opt} gives the local and global consistency error function~\cite{zhou2004learning},
\begin{align}
  \mathbb{E}(F)=\frac{1}{2}\sum_{i,j}{W_{ij} \sum_c{\left\| \frac{F_{ic}}{\sqrt{D_{ii}}} - \frac{F_{jc}}{\sqrt{D_{jj}}} \right\|^2 }} \notag \\
  + \left( \frac{1}{\alpha}-1 \right) \sum_i\sum_c{ \left\| F_{ic} - B_{ic} \right\|^2} \enspace.
  \label{eq_costLP}
\end{align}
Equation~\eqref{eq_costLP} can be minimised using a closed form expression. However, since this expression involves a matrix inversion and $\mathcal{G}_s$ is usually constructed to be sparse (either by using an exponentially weighted similarity function or only connecting the $k$-nearest neighbours~\cite{belkin2004semi, talukdar2009new, zhou2004learning, zhu2003semi}), it is preferable to avoid this computational and memory expensive operation by using the power method to iteratively update $\mathbf{F}$: 
%
%
\begin{equation}
  \mathbf{F}_{t+1} = \mathbf{Z}^{-1}((1-\alpha)\mathbf{B} + \alpha\mathbf{LF}_{t}) \enspace,
  \label{eq_lp}
\end{equation}
where $\mathbf{Z}^{-1}$ is a diagonal matrix used to normalise the rows of $\mathbf{F}$.  Equation~\ref{eq_lp} is often referred to as label propagation since the term $\mathbf{LF}$ propagates label information from each node to its neighbours. The parameter $0\leq \alpha \leq 1$ controls the rate at which the label information is propagated around the network. Once the algorithm has converged, we can make predictions according to the maximum label scores $y_i^* = \arg \max_c \; F_{ic}$.
  
Na\"{\i}vely, we can apply this method to relational networks but 
 the fundamental assumption that labels vary smoothly across the graph means that 
 the approach fails when the labels are not assortative~\cite{peel2011topological}.

\subsection{Heterogeneity in relational networks}
\label{sec_hetero}

Relational networks are not constructed based on the similarities of the nodes, but instead links may form because of similarities \textit{or} differences (or combination thereof) in node attributes.  Also, there may be local patterns of similarity or difference between connected nodes, or there may be no relationship at all between node attributes and network structure~\cite{hric2016network, peel2016ground}. 
  Here we describe these concepts by the heterogeneity of the network, specifically: \textit{link-heterogeneity} and \textit{class-heterogeneity}.

  Link-heterogeneity (also known as disassortativity or heterophily) is when links are more likely between nodes of a different class than between those of the same.  Bipartite and multi-partite networks are strict forms link-heterogeneity, while core-periphery, hierarchical and mixtures of these structures are cases of partial link-heterogeneity.  Link heterogeneity makes the node classification problem difficult as we need to determine the pattern of interactions between classes, which may easily be obscured by the fact that the network is only partially labelled.  
 The fewer explicitly labelled nodes we have available, the harder it becomes to determine the pattern of links.   

Class-heterogeneity is when nodes of the same class exhibit different patterns of links.  For example, in Figure~\ref{fig_links}(c) we have two classes of nodes ($\{red, black\}$) and half of each class is assortative while the other half is disassortative.  In other words, if $\mathcal{C}$ is a set of nodes belonging to a single class, then there may exist subsets $\mathcal{P},\mathcal{Q} \subset \mathcal{C}$ such that $\mathcal{P}$ is assortative and $\mathcal{Q}$ is disassortative.    
 If one of the patterns of class interactions appears more frequently than another, then we may only discover the dominant pattern and miss the others, e.g.~if a class has more assortative links than disassortative links, then we might assume that all links are assortative and incorrectly classify nodes with disassortative links.


\section{Similarity in networks}
\label{sec_similarity_in_networks}
A key step in GSSL is constructing a graph based on similarity, so to apply GSSL to relational networks we should first transform the relational network into a similarity graph. To do this, we require some notion of similarity between network nodes based on the pattern of links that is robust to the aforementioned heterogeneities.  Here, we consider node equivalences from social network theory as a type of similarity. 
Regular equivalence has long been used in social network analysis to identify nodes that play a similar role in the network~\cite{Borgatti1989class,White1983Graph}. 
\begin{definition}{Regular equivalence}~\cite{White1983Graph} ---\\
 If $\cong$ is an equivalence relation on $\mathcal{V}_r$ then $\cong$ is a regular equivalence iff $\forall a,b,c \in \mathcal{V}_r$ and  $a\cong b$:
 \begin{enumerate}
   \item $\forall (a \rightarrow c) \in \mathcal{E}_r, \exists d \in \mathcal{V}_r$ s.t. $(b \rightarrow d) \in \mathcal{E}_r$ and $c \cong d$;
   \item $\forall (a \leftarrow c) \in \mathcal{E}_r, \exists d \in \mathcal{V}_r$ s.t. $(b \leftarrow d) \in \mathcal{E}_r$ and $c \cong d$.
 \end{enumerate}
 \label{def_regequiv}
\end{definition}
A stochastic variant of regular equivalence, in which the existence of an edge is replaced with the same probability of an edge $\textrm{Pr} \!\left[(a \rightarrow c) \in \mathcal{E}_r\right]$, is the foundation of the popular generative network model called the stochastic blockmodel (SBM)~\cite{holland1983stochastic,nowicki2001estimation}.  The SBM is frequently used to detect communities that are assortative~\cite{karrer2011stochastic}, disassortative~\cite{Larremore2014Efficiently} or a mixture of the two~\cite{Zhang2015Identification}.  The SBM has previously been used for semi-supervised network node classification~\cite{zhang2014phase, gatterbauer2015linearized, Moore2011Active} and works well under link-heterogeneity.  However, because the SBM treats all nodes of the same class as stochastically equivalent it is unable to deal with class-heterogeniety, as we will show in Sec.~\ref{sec_results}. 
 Extensions of the SBM have been shown to perform favourably on the node classification problem~\cite{peel2011topological,peel2012supervised}, but the computational expensive of these methods means they do not scale well to large networks.  Our goal is to use node equivalence in the GSSL framework to gain the flexibility of the SBM and the scalability of GSSL.

Definition~\ref{def_regequiv} tells us that for nodes to be regularly equivalent there is no requirement for them to be connected by a link, only that their neighbours are equivalent.  The recursive nature of this definition means, however, that we cannot calculate it directly. Instead, we approximate regular equivalence through the relaxation of a stricter equivalence formed by setting $c=d$.
\begin{definition}{Structural equivalence}~\cite{lorrain1971structural} ---\\
 If $\equiv$ is an equivalence relation on $\mathcal{V}_r$ then $\equiv$ is a structural equivalence iff $\forall a,b,c \in \mathcal{V}_r$ and  $a\equiv b$:
 \begin{enumerate}
   \item $\forall (a \rightarrow c) \in \mathcal{E}_r$, then $(b \rightarrow c) \in \mathcal{E}_r$;
   \item $\forall (a \leftarrow c) \in \mathcal{E}_r$, then $(b \leftarrow c) \in \mathcal{E}_r$.
 \end{enumerate}
 \label{def_structequiv}
\end{definition}
Usually, in real networks, finding sets of nodes with exact structural equivalence is rare, so instead we consider two different relaxations of structural equivalence, each of which will form the basis for one of our two proposed methods.  

\subsection{Common neighbours} 
\label{sec_common} 
In place of identifing nodes with exact structural equivalence, we can compare neighbour sets using cosine similarity~\cite{Salton1989Automatic}, for which $0$ indicates no common neighbours and $1$ is structural equivalence. 
Cosine similarity $S_{a,b}$ between nodes $a$ and $b$ is:
\begin{equation}
  S_{a,b} = \frac{\sum_{c}{A_{ac}A_{cb}}}{\sqrt{\sum_{c}{A_{ac}^2}} \sqrt{\sum_{c}{A_{cb}^2}} } \enspace,
  \label{eq_cos_sim}
\end{equation}
where $A_{ab}$ is the 
adjacency matrix entry at row $a$ and column $b$. 
For an undirected network the ajacency matrix is symmetric, so in matrix form the undirected similarity $\mathbf{S}_{\leftrightarrow}$  is:
\begin{equation}
  \mathbf{S}_{\leftrightarrow} = \mathbf{D}^{-\frac{1}{2}}\mathbf{A}\mathbf{A}\mathbf{D}^{-\frac{1}{2}}\enspace,
  \label{eq_cos_sim_mat}
\end{equation}
where $\mathbf{D}$ is a diagonal matrix of node degrees.  If the network is directed, then we can also calculate directed similarity based on the common neighbours linked-to ($S_{\rightarrow}$) and linked-from ($S_{\leftarrow}$),
%
%
\begin{equation}
  \mathbf{S}_{\rightarrow} = \mathbf{D}_{\rightarrow}^{-\frac{1}{2}}\mathbf{A}\mathbf{A^T}\mathbf{D}_{\rightarrow}^{-\frac{1}{2}} \, , 
  \;
  \mathbf{S}_{\leftarrow} = \mathbf{D}_{\leftarrow}^{-\frac{1}{2}}\mathbf{A^T}\mathbf{A}\mathbf{D}_{\leftarrow}^{-\frac{1}{2}}\, ,
  \label{eq_cos_sim_mat2}
\end{equation}
where $\mathbf{D}_{\rightarrow}$ and $\mathbf{D}_{\leftarrow}$ are the diagonal matrices of out-degree and in-degree respectively.

\subsection{Neighbours of neighbours}
\label{sec_neigh}
Definition~\ref{def_structequiv} implies that for an undirected network (or a directed network in which we ignore link direction) there will always be a path of length two between structurally equivalent nodes.  That is, if $a \equiv b$ and $(a\leftrightarrow c) \in \mathcal{E}$, then $(c\leftrightarrow b) \in \mathcal{E}$.
\begin{theorem} {(Neighbours of neighbours)}
The set of neighbours of a node's neighbours contains all of the nodes structurally equivalent to it. 
\end{theorem}
If $\mathcal{H}_b$ is the set of nodes that are structurally equivalent to $b$ and $\mathcal{M}_b$ is the set of all neighbours of $b$'s neighbours, then $\mathcal{H}_b \subseteq \mathcal{M}_b$.  We can therefore use the set $\mathcal{M}_b$ as a set of candidate nodes for structural equivalence with $b$.

\section{Label propagation in relational networks}
\label{sec_methods}
Label propagation, as described in Sec.~\ref{sec_gssl}, allows for class-heterogeneity through the cluster assumption.  This is the assumption that clustered instances will have the same class label, but it does not constrain all instances of a class to be in the same cluster.  
The similarity measures of the previous section allow us to identify nodes with similar patterns of links.  
By performing label propagation on similarity graphs based on these  measures, we should be able to perform node classification for relational networks with link-heterogeneity and/or class-heterogeneity.  


In what follows, we describe our two new methods \textit{cosine label propagation} and \textit{two-step label propagation} based on the common neighbours (Sec.~\ref{sec_common}) and neighbours of neighbours (Sec.~\ref{sec_neigh}) similarities respectively.  

\subsection{Cosine label propagation}
\label{sec_cosine_label}
We can transform the relational network into a similarity graph with edge weights corresponding to the number of common neighbours each pair of nodes shares.  For an undirected network, the adjacency of this similarity graph $\mathbf{A}_s$ is given by adjacency matrix of the relational network squared, i.e.~$\mathbf{A}_s =\mathbf{A}_r\mathbf{A}_r$.  Noticing that the cosine similarity of the relational network takes the form of the normalised laplacian of the common neighbours similiarity graph (but normalised by the degree sequence of the relational network -- Eq.~\eqref{eq_cos_sim_mat}), we substitute the similarity matrix $\mathbf{S}_{\leftrightarrow}$ in place of $\mathbf{L}$ in Eq.~\eqref{eq_lp}, i.e.,
%
%
 \begin{equation}
   \mathbf{F}_{t+1} = \mathbf{Z}^{-1}\left((1-\alpha)\mathbf{B} + \alpha \mathbf{S}_{\leftrightarrow}\mathbf{F}_{t} \right) \enspace.
   \label{eq_mani}
 \end{equation}
For directed networks we can similarly use the directed similarity matrices in Eq.~\eqref{eq_cos_sim_mat2}.
  However, this means we have to make a choice about which of the similarity matrices to use.  Furthermore, these similarity matrices may be much denser than the original adjacency matrix and so their explicit construction can cause memory issues for large networks.  We can solve both these problems using a low rank approximation of $\mathbf{S}$ based on the eigenvectors of $\mathbf{S}$ with the largest eigenvalues. This is a useful approximation because these eigenvectors represent smooth functions on a graph~\cite{zhu2004nonparametric} and because we are working with a similarity graph, it becomes reasonable to once again assume smoothness. These eigenvectors can be efficiently computed (e.g. using the power method) from the singular value decomposition of $\mathbf{D}^{-\frac{1}{2}}\mathbf{A}$ without needing to explicitly construct $\mathbf{S}$. Furthermore, we can combine features using the top $k$ eigenvectors from each of the different $\mathbf{S}$ matrices.  We can then use the $n \times vk$ matrix (where $v$ is the number of $\mathbf{S}$ matrices) of eigenvectors $\mathbf{\Phi}$ for label propagation:
 \begin{equation}
   \boxed{\mathbf{F}_{t+1} = \mathbf{Z}^{-1}\left((1-\alpha)\mathbf{B} + \alpha \mathbf{\Phi}\mathbf{\Phi^T}\mathbf{F}_{t} \right)} \enspace,
   \label{eq_cos_lp}
 \end{equation}
which we iterate until convergence.  


 
 
%

\subsection{Two-step label propagation}
\label{sec_2step}
Label propagation passes label information across links based on the assumption that connected nodes are more likely to be the same class.  This may not be the case for relational networks, so instead we propose to propagate label information to nodes that are two steps away by skipping the immediate neighbours, i.e.~propagate labels to the neighbours of their neigbours:
%
%
\begin{equation}
  \boxed{\mathbf{F}_{t+1} = \mathbf{Z}^{-1}((1-\alpha)\mathbf{B} + \alpha\mathbf{(LL)^\beta F}_{t})} \enspace,
  \label{eq_2step}
\end{equation}
where $\beta$ is a parameter that allows for taking multiples of two steps at a time.  This two-step label propagation continues the notion that structural equivalence implies similarity rather than connectivity.  We know that there always exists a path of length two between structurally equivalent nodes (see Sec.~\ref{sec_neigh}).  Although a path of length two does not imply either regular or structural equivalence, nodes with more paths of length two between them are closer to being equivalent.  Consequently, the more paths of length two between nodes, the more they influence each others label prediction.  

An alternative view is that label propagation relates to solving an eigenvector problem. 
While the eigenvectors of $\mathbf{L}$ and $\mathbf{LL}$ are the same, the eigenvalues are different. The eigenvalues of the latter are the square of the former.  Therefore, negative eigenvalues of $\mathbf{L}$ become positive in $\mathbf{LL}$ and the order of the dominant eigenvectors changes such that the dominant eigenvectors correspond to the most and least smooth functions over the graph. The change to the spectrum gives the two-step label propagation algorithm the opportunity to capture disassortative as well as assortative patterns in the network.  The parameter $\beta$ also modifies the spectrum, which we demonstrate to be beneficial in some cases (Sec.~\ref{sec_results}).

\subsection{Time complexity}
The label propagation update equation (Eq.~\eqref{eq_lp}) comprises a matrix-vector multiplication, a matrix addition and  a normalisation step.  As $\mathbf{L}$ is sparse, with $m$ nonzero entries, the multiplication $\mathbf{LF}$ has time complexity $O(m\ell)$.  Both the matrix addition $\mathbf{B} + \alpha\mathbf{LF}$ and normalisation step can be computed in $O(n\ell)$.  Therefore the time complexity of each label propagation iteration is $O(n\ell + m\ell)$.  
Since both proposed methods are forms of label propagation, their time complexity is comparable.  For cosine label propagation (Eq.~\eqref{eq_cos_lp}), the matrix $\mathbf{\Phi}$ tends to be dense containing $m=nvk$ nonzero entries, so each iteration is $O(n\ell + nvk\ell)$. 
Two-step label propagation (Eq.~\eqref{eq_2step}) has the same time complexity as regular label propagation as the ``extra step'' can be calculated by multiplying $\mathbf{L}$ by $\mathbf{LF}$ in $O(m\ell)$.  


\section{Experimental Results}
\label{sec_results}
We compare the performance of the following methods on a variety of synthetic and real-world networks 
 {\color{blue}(Our new methods in blue)}:
\begin{itemize}[noitemsep,topsep=6pt,leftmargin=*]
  \item \textit{1-step LP}~\cite{zhou2004learning}: Regular label propagation. 
  \item \textit{linBP}~\cite{gatterbauer2015linearized}: Linearised belief propagation is a state-of-the-art approach that allows for link-heterogeneity by means of an $\ell \times \ell$ affinity matrix, $\mathbf{H}$, that indicates the relative propensity of links between classes in a similar manner as the SBM.  
  It is an iterative algorithm that updates according to $\mathbf{\hat{F}}_{t+1} = \mathbf{\hat{B}} + \mathbf{A\hat{F}}_{t}\mathbf{H} - \mathbf{D\hat{F}}_{t}\mathbf{H}^2$ .
  %
  %
  \item \textit{linBP I}: Same as \textit{linBP} above, but with $\mathbf{H}$ replaced by the identity matrix $\mathbf{I}$. 
  \item \textit{ghost}~\cite{gallagher2008using}: Similar to our methods, it also transforms the network and considers even-length paths between nodes.  
``Ghost'' edges are inserted to link every unlabelled node to each labelled node with edge weights that are  scores of an even-step random walk with restart starting from each labelled node. 
  \item \textit{mmsb}~\cite{peel2011topological}: An extended mixed membership stochastic block model to include a distribution over class labels conditioned on block membership.  
  \item \textit{\color{blue} cosine undirected}: Cosine label propagation (Sec.~\ref{sec_cosine_label}) using the top $k$-eigenvectors of $\mathbf{S}_{\leftrightarrow}$.
  \item \textit{\color{blue} cosine directed}: Cosine label propagation using the top $k$-eigenvectors of $\mathbf{S}_{\leftarrow}$ and $\mathbf{S}_{\rightarrow}$.
  \item \textit{\color{blue} cosine both}: Cosine label propagation using the top $k$-eigenvectors of $\mathbf{S}_{\leftrightarrow},\mathbf{S}_{\leftarrow}$ and $\mathbf{S}_{\rightarrow}$.
  \item \textit{\color{blue} 2-step LP}: Two-step label propagation as described in Section~\ref{sec_2step}.
\end{itemize}

We set the parameters $\{k,\alpha, \beta \}$ using 3-fold cross validation and selected from $k=\{5,15,25\}, \alpha=\{0.01,0.1,0.5,0.9\}$ and $\beta=\{1,2,3\}$.  Sensitivity to different parameter settings is discusssed in Section~\ref{sec_params}.

\begin{figure}
  \includegraphics[width=\columnwidth]{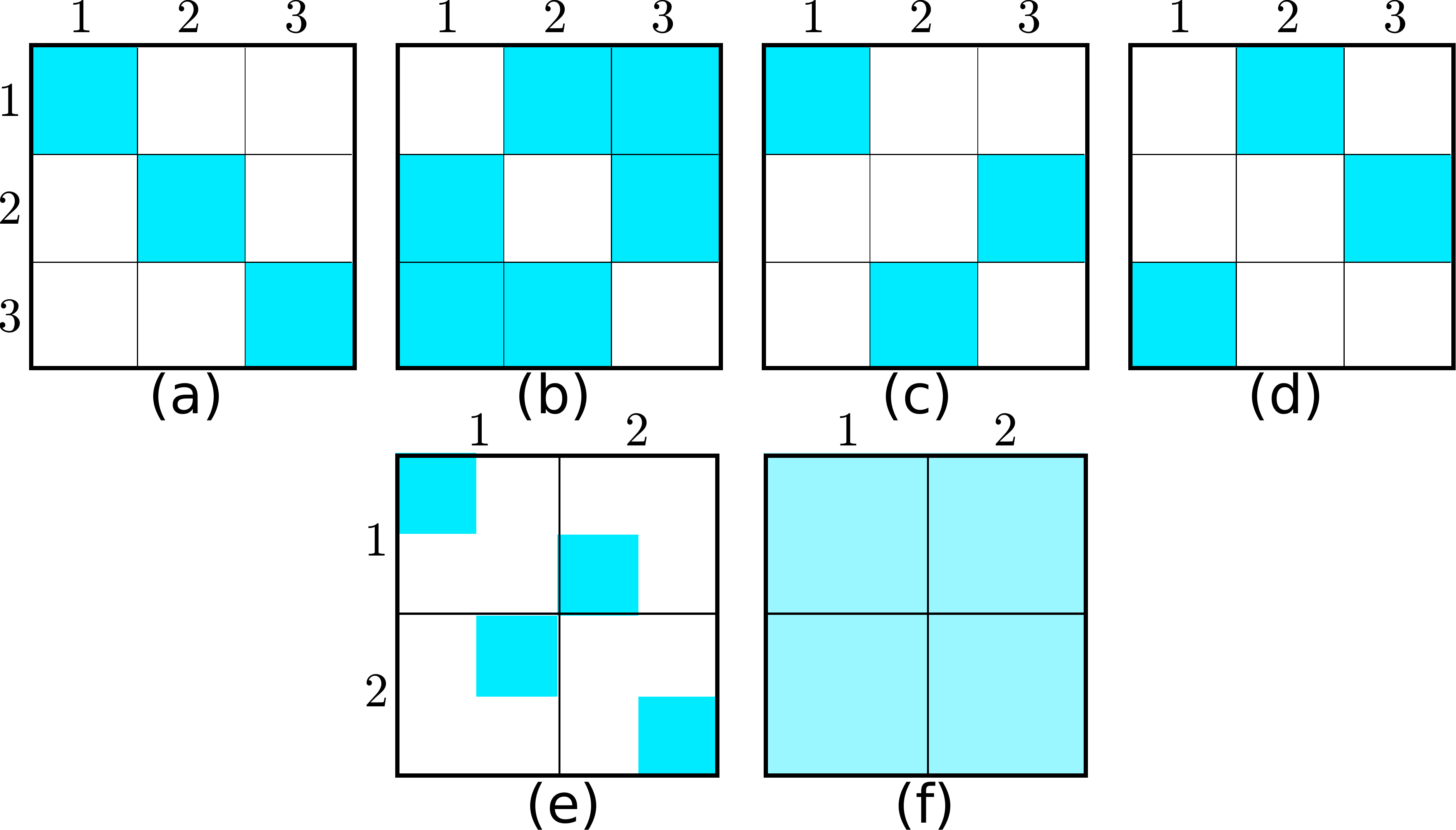}
  \caption{Block interaction matrices used to generate synthetic networks.  White ($p_w$) and blue ($p_b$) blocks represent different link probabilities; $p_b>p_w$. (a) 3 assortative classes, (b) 3 disassortative classes, (c) 3 mixed classes, (d) 3 cyclic classes, (e) 2 heterogeneous classes. 
  (f) Link probabilities appear equal when treating the classes in (e) as homogeneous.
  }
  \label{fig_synthblocks}
\end{figure}

\subsection{Synthetic networks}
We first test our methods on synthetically generated networks in which we can control the network structure and its relationship to the class labels.   We use an SBM 
to generate the network and node labels.  The SBM is a probabilistic generative network model based on stochastic equivalence.  

The SBM assigns each node to one of $\kappa$ groups.  Links are generated between nodes conditional on the groups they are assigned to and a $\kappa \times \kappa$ affinity matrix, $\omega$, such that $P(A_{ij}=1| g_i, g_j, \omega) = \omega_{g_i,g_j}$, 
where $g_i$ indicates the group assignment of node $i$.  The node label $y_i$ is then assigned according to a deterministic mapping $h(g): g \rightarrow y$. This model is able to generate networks with both types of heterogeneity described in Section~\ref{sec_hetero}.  Link-heterogeneity (disassortativity) is generated by large off-diagonal elements in $\omega$ and class-heterogeneity is generated by mapping multiple groups to a single class label.  

Figure~\ref{fig_synthblocks}(a)--(e) shows a graphical representation of the interaction matrices $\omega$ used to generate the synthetic networks.  We parameterise $\omega$ using two parameters $p_b$ and $p_w$, corresponding to the blue and white blocks respectively.  In the first four cases there are three groups each mapped to a single class label, i.e.~$\kappa = \ell = 3$ and $h(g)$ is a one-to-one mapping.  Each case corresponds to a different type of interaction: (a) \textit{assortative} -- nodes of the same group prefer to link to each other, (b) \textit{disassortative} -- nodes prefer to link to nodes from different groups, (c) \textit{mixed} -- one assortative group and two disassortative groups, and (d) \textit{cyclic} -- each group prefers linking to a different group, i.e.~$g_1 \rightarrow g_2$, $g_2 \rightarrow g_3$ and $g_3 \rightarrow g_1$.  In the fifth case (e) \textit{heterogeneous} -- we have four groups mapped to two class labels ($\kappa=4,\ell=2$) to simulate class-heterogeneity such that half of each class is assortative and the other half is disassortative.  

\begin{figure}
  \centering
  \includegraphics[width=\columnwidth]{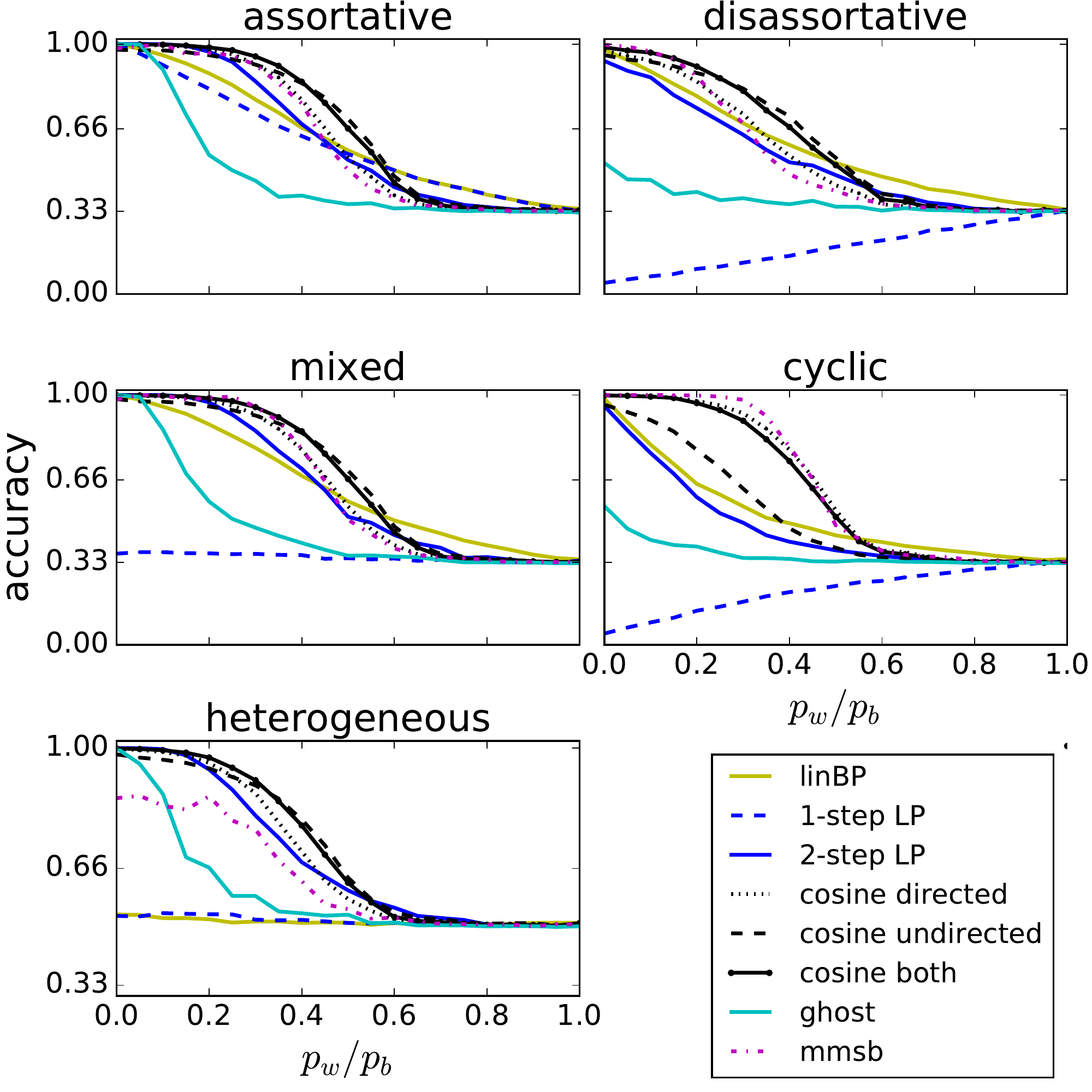}
  \caption{Classification accuracy on synthetic networks (see Fig.~\ref{fig_synthblocks}) as a function of $p_w/p_b$.  Our methods give comparable or better performance than state-of-the-art baseline models.}
  \label{fig_synthacc}
\end{figure}

Figure~\ref{fig_synthacc} shows the accuracy of predicting the class labels of the unlabelled nodes as we decrease the strength of the block structure by increasing the ratio $p_w/p_b$ from $0$ (strong structure, no links in the white blocks) to $1$ (no structure, equal probability of links in all blocks).  In all experiments we use the same number of nodes $n=10^3$, of which 10\% are labelled ($|\mathcal{L}|/n=0.1$), and a constant mean node degree of $15$.  We see that for high values of $p_w/p_b$ ($>0.7$) all methods perform poorly, presenting little or no improvement over 
random labelling (i.e.~$1/\ell$).  This is because as $p_w/p_b \rightarrow 1$ 
the networks become indistinguishable from random graphs~\cite{Decelleetal2011, zhang2014phase}. 

For lower values of $p_w/p_b$, we see that our methods perform comparably or better than the baseline algorithm \textit{linBP}.  In particular \textit{linBP} performs badly in the \textit{heterogeneous} experiments since it assumes that all nodes with a particular class label link to the rest of the network in the same way. Under this assumption, nodes of both classes appear the same (Fig.~\ref{fig_synthblocks}(f)).  In all cases our methods outperform regular label propagation and the ghost edges method.

Overall the \textit{cosine} methods perform best. The \textit{cosine undirected} had the highest accuracy in all but the \textit{mixed} experiments, where the \textit{cosine directed} performed much better.  However, \textit{cosine both} method came a close second in every experiment, providing the ``best of both worlds''.  We therefore focus only on the \textit{cosine both} in the experiments on real networks.

\subsection{Real networks}
\begin{figure*}
  \centering
  \includegraphics[width=\textwidth]{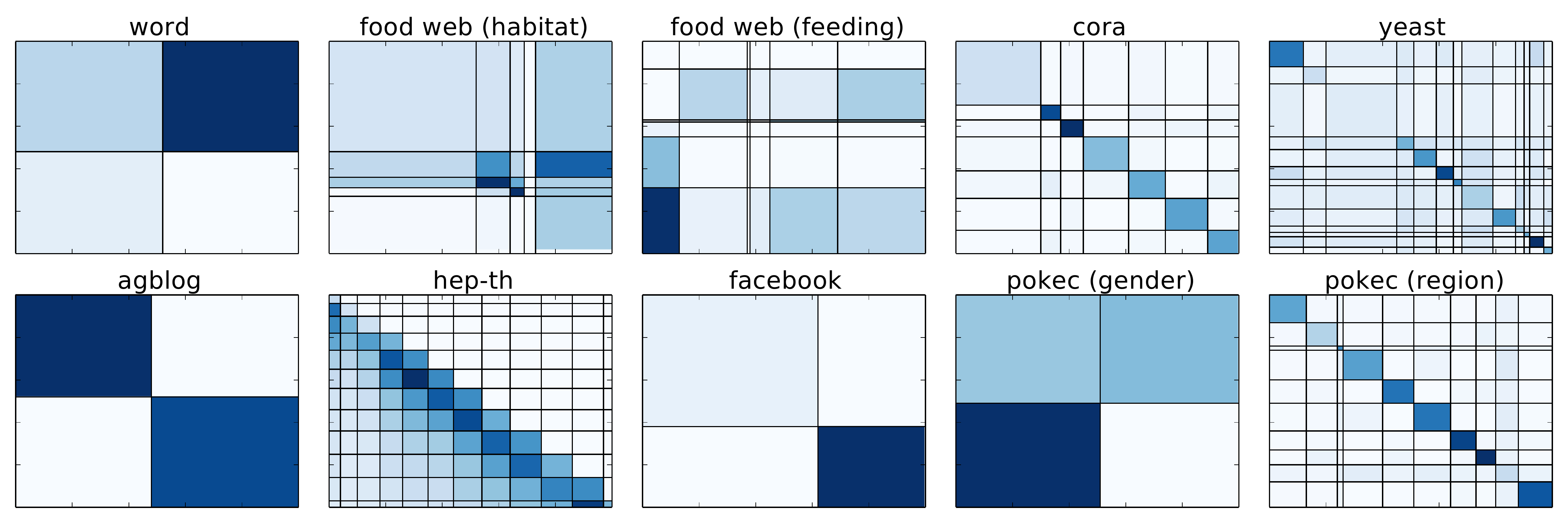}
  \caption{An illustration of relative link densities (darker blue indicates higher density) within and between classes in the real world networks.  Row and column sizes are proportional to class size.}
  \label{fig_realblocks}
\end{figure*}

We now compare the performance on a variety of publicly available real-world networks with different size and structure.  These networks span a broad range of domains, including language (\textit{word}), ecological (\textit{foodweb}), citation (\textit{cora, hep-th}), web (\textit{blog}), and social (\textit{facebook, pokec}) networks.  Table~\ref{tab_networks} provides a summary of the number of nodes ($n$), edges ($m$) and class labels ($\ell$) in each network.  In some networks there are nodes missing their class labels.  
We exclude these nodes when assessing the performance, but keep them in the network to preserve the structure. 
In Table~\ref{tab_networks}, $\tilde{n}$ is the number of nodes with missing labels.  Figure~\ref{fig_realblocks} illustrates the network structure with respect to the relative density of links within and between classes; darker blue indicates higher densities.  The relative size of the rows and columns indicates the proportion of nodes in each class.  

\begin{table}
\centering
\caption{Real network datasets }
  \label{tab_networks}
  \begin{tabular}{|l|c|c|c|c|}
  \hline
    Network (label) & $n$ & $\tilde{n}$ & $m$ & $\ell$ \\
    \hline
    word (adj/noun)~\cite{newman2006finding} & 112 & 0 & 569 & 2\\
    foodweb (habitat)~\cite{brose2005body} & 492 & 4 & 16330 & 5\\
    foodweb (feeding)~\cite{brose2005body} & 492 & 4 & 16330 & 6\\
    cora (subject)~\cite{sen2008collective} & 2708 & 0 & 5429 & 7\\
    yeast (function)~\cite{bu2003topological} & 2361 & 0 & 7182 & 13\\  
    agblog (political)~\cite{adamic05thepolitical} & 1222 & 0 & 33428 & 2\\   
    hep-th (year)~\cite{gehrke2003overview} & 27770 & 0 & 352807 & 12\\   
    facebook (gender)~\cite{mcauley2012learning} & 4039 & 0 & 176468 & 2\\   
    pokec (gender)~\cite{takac2012data} & 1632803 & 163 & 30622564 & 2\\   
    pokec (region)~\cite{takac2012data} & 1632803 & 163 & 30622564 & 10\\   
    \hline
  \end{tabular}
\end{table}

\begin{figure*}
  \centering
  \includegraphics[width=\textwidth]{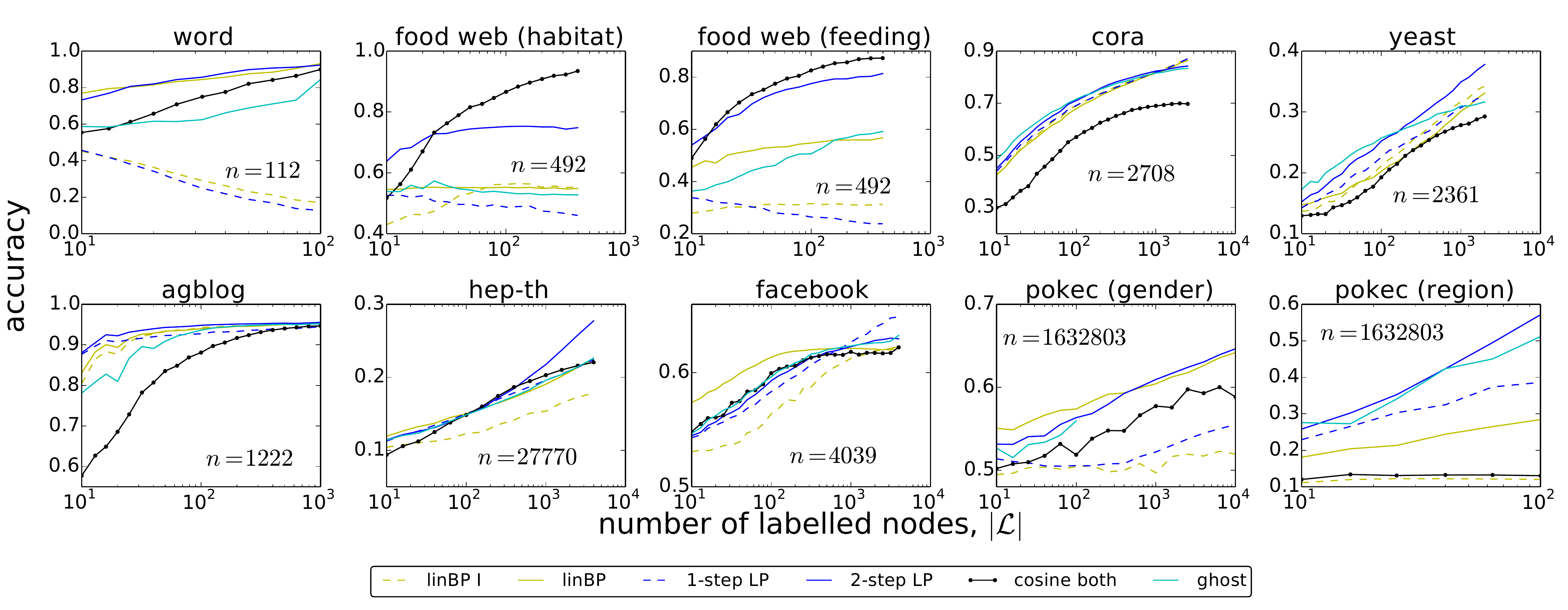}
  \caption{Classification accuracy (\textit{y-axis}) on real-world networks as the size of the labelled set (\textit{x-axis}) is varied. The \textit{2-step LP} method has the best overall accuracy.}
  \label{fig_realacc}
\end{figure*}

Figure~\ref{fig_realacc} shows the classification accuracy on the real networks as we vary the number of nodes in the labelled set $\mathcal{L}$. We see that the baselines that assume assortativity, \textit{linBP I} and \textit{1-step LP}, perform poorly on networks that display disassortative connectivity (Fig.~\ref{fig_realblocks}).  The \textit{ghost} method does not perform particularly well overall, only slightly outperforming the other methods on \textit{cora} and \textit{yeast} when the training set is very small.

Compared to \textit{linBP}, at least one of our methods outperforms it on all except the  \textit{facebook} and \textit{pokec (gender)} networks.  In those cases \textit{linBP} performs better only when the training set is small.  It is important to realise that we provide the \textit{linBP} method with complete knowledge of the class interactions (by way of the affinity matrix $\mathbf{H}$) and so it knows \textit{a priori} if the network is assortative or disassortative. In contrast, our methods do not. Recent work~\cite{gatterbauer2014Semi} investigates methods for estimating $\mathbf{H}$ from the data but, as we demonstrate here, even with complete information about class affinities, \textit{linBP} is often unable to outperform our \textit{2-step} label propagation. In most cases the \textit{2-step LP} gives the best or second-best performance.  The \textit{cosine} method, however, is less consistent, ranging from best (\textit{food web}) to worst (\textit{agblog, cora, yeast}) performance.  


\begin{figure*}
  \centering
  \includegraphics[width=\textwidth]{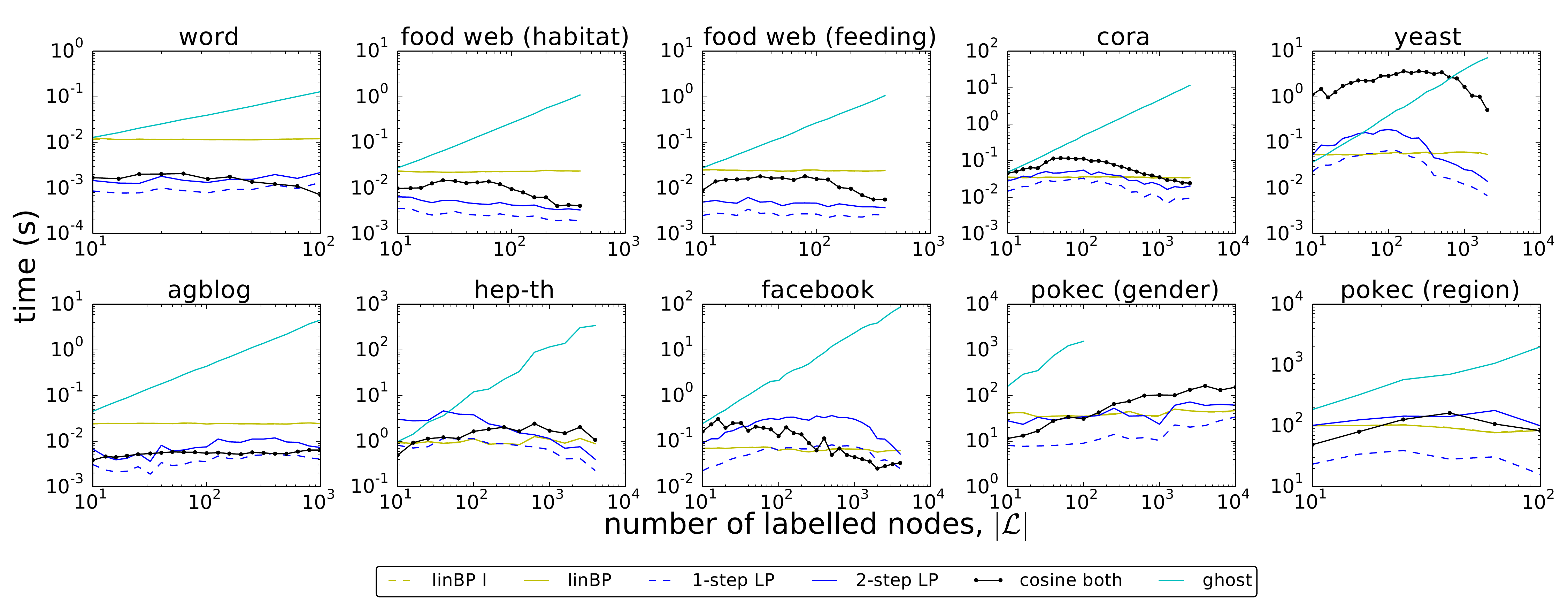}
  \caption{Average run times on real world networks as a function of number of labelled nodes.} 
  \label{fig_realtime}
\end{figure*}

Figure~\ref{fig_realtime} shows the average run times for each method on each of the networks as a function of training set size.
We see that the ghost method scales with number of labelled nodes --- for large networks like pokec it took hours to run when the number of labelled nodes was $O(10^2)$. In contrast, the run times of our methods is relatively constant as the size of the training set varies. In some cases we even observe a decrease in time for larger training sets as less iterations are needed for the algorithm to converge.

Figure~\ref{fig_realtime} shows the average run times for each method on the policital blogs network as a function of training set size. We see that the \textit{ghost} method scales with number of labelled nodes --- for large networks like \textit{pokec} it took hours to run when the number of labelled nodes was $O(10^2)$.  In contrast, our methods 
are relatively constant as the size of the training set varies.  

\begin{figure*}[t]
  \centering
  \includegraphics[width=\textwidth]{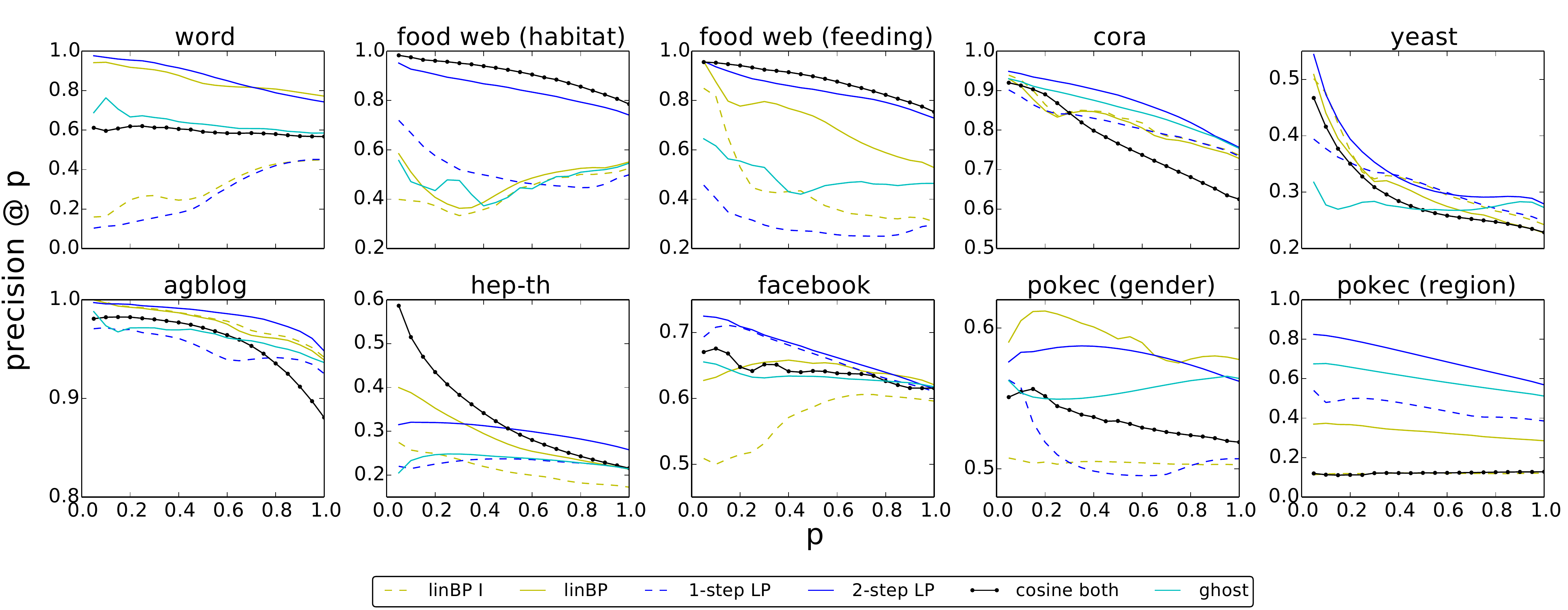}
  \caption{Precision@$p$ (\textit{y-axis}) on real-world networks as a function of the proportion ``$p$'' (\textit{x-axis}) of the unlabelled nodes with the highest label score $F$.   The \textit{2-step LP} method tends to produce a monotonically decreasing precision with increasing $p$, suggesting that the label score $F$ of \textit{2-step LP} is a reliable measure of confidence.} 
  \label{fig_realpatp}
\end{figure*}

In Figure~\ref{fig_realpatp} we see the precision@$p$ for each of the methods applied to the real networks.  In all cases, except for the \textit{pokec} network, $10\%$ of the nodes were labelled with the remaining $90\%$ used as the test set. For the \textit{pokec} network we set the training set size $|\mathcal{L}|=100$ so that we could compare against the \textit{ghost} baseline.  The precision@$p$ is the precision of the top $p$ proportion of unlabelled nodes ordered by their maximum label score $\max_c F_{ic}$.  When $p=1$ the precision@$p$ is the precision on all of the unlabelled nodes and is equal to the accuracy.  Overall we see the best precision is obtained with the \textit{2-step LP} method.  Furthermore, both our methods, \textit{2-step LP} and \textit{cosine}, (with the exception of \textit{pokec}) tend to decrease as $p$ increases.  This result suggests that the \textit{2-step LP} label score is a reliable measure of confidence in the predicted label and so could be useful for tasks such as active learning~\cite{Moore2011Active, Peel2015Active}, which often involves estimating the uncertainty of predictions.

\subsection{Parameter settings}
\label{sec_params}
\begin{figure*}[t]
  \centering
  \includegraphics[width=\textwidth]{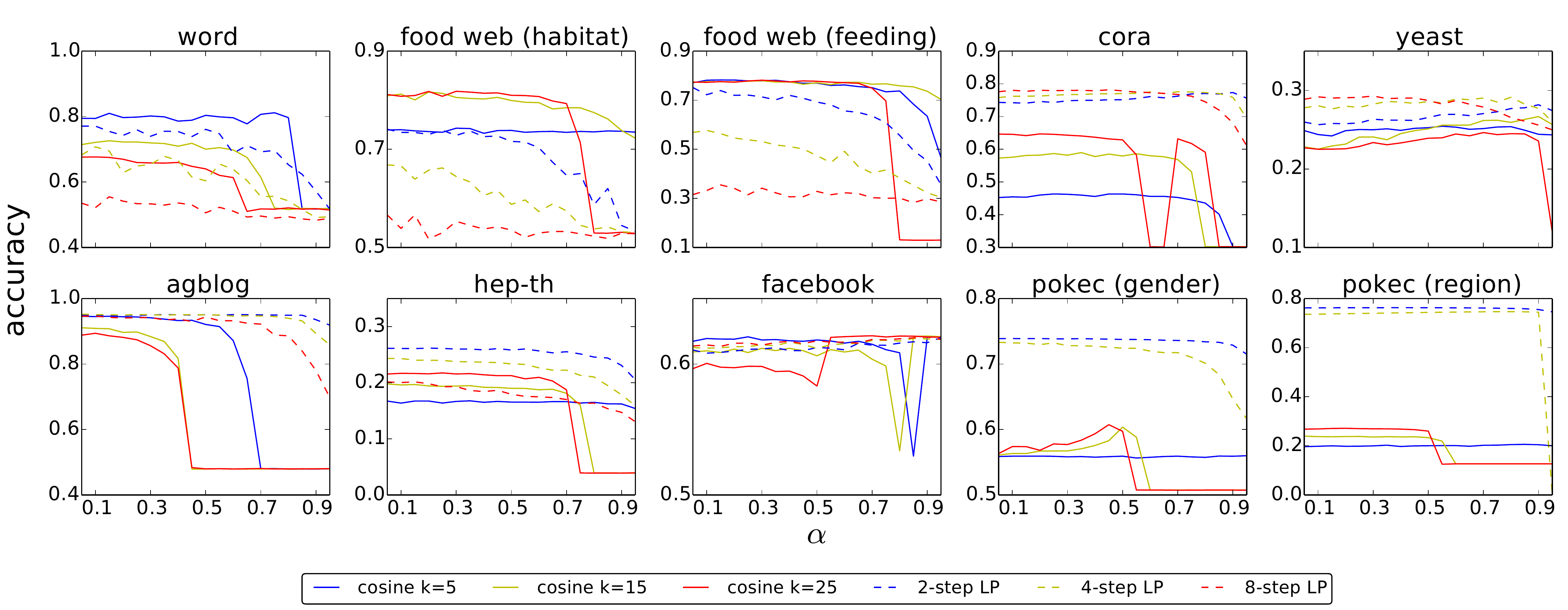}
  \caption{The accuracy  (\textit{y-axis}) for different values of $\alpha$ (\textit{x-axis}).  The solid lines show the \textit{cosine} method for different values of $k$ and the dashed line shows \textit{2-step LP}.  The \textit{2-step LP} method has only a single parameter and consistently performs well for a wide range of values.}
  \label{fig_params}
\end{figure*}

Throughout the experiments we chose the parameters using cross-validation.  In this section we discuss the effect different parameter settings have on these approaches.  
We repeated the experiments on real networks for different parameter values for $\alpha, k$(\textit{cosine} method) and  $\beta$ (\textit{2-step LP} method). 
Figure~\ref{fig_params} shows the accuracy on the y-axis against the parameter $\alpha$ on the x-axis.  The solid lines show the \textit{cosine} method for different values of $k$ and the dashed line shows \textit{2-step LP} for different values of $\beta$.   

For the \textit{cosine} method, higher values of $k$ often improve accuracy but, as the algorithm scales with $k$, this comes at an increased computational cost (see Table~\ref{tab_runts}).  On some networks, however, lower values of $k$ are preferable.  In general low values of $\alpha$ are usually better, but in some cases e.g.~pokec (gender), mid-range values of $\alpha$ are optimal for certain values of $k$.  
For the \textit{2-step LP} method, lower values of $\beta$ are usually best. The method performs consistently well for a wide range of $\alpha$ values.  We see from Table~\ref{tab_runts} that lower values of $\alpha$ result in a faster run time, which is due to the algorithm converging in a lower number of iterations.  For our experiments of $\alpha=0.1$, the algorithm would converge in under 10 iterations.

\begin{table}
\small
\centering
\setlength\tabcolsep{4.5pt}
\caption{Run times (s)}
  \label{tab_runts}
  \begin{tabular}{|l r|c|c|c|c|c|c|}
  \hline
     \multicolumn{2}{|l}{} & \multicolumn{4}{|c|}{\textit{cosine}} & \multicolumn{2}{c|}{\textit{2-step LP}}\\
     & $k$ &\multicolumn{2}{|c|}{$5$} & \multicolumn{2}{c|}{$15$} & \multicolumn{2}{c|}{--} \\
     & $\alpha$ & $0.1$ & $0.6$ & $0.1$ & $0.6$ & $0.1$ & $0.6$ \\
    \hline
    \multicolumn{2}{|l|}{word} & 0.010 & 0.012 & 0.016 & 0.019 & 0.001 & 0.002\\
    \multicolumn{2}{|l|}{foodweb (h)}& 0.019 & 0.023 & 0.026 & 0.035 & 0.003 & 0.011 \\
    \multicolumn{2}{|l|}{foodweb (f)}& 0.016 & 0.022 & 0.025 & 0.036 & 0.003 & 0.014\\
    \multicolumn{2}{|l|}{cora} & 0.070 & 0.097 & 0.194 & 0.248 & 0.005 & 0.017\\
    \multicolumn{2}{|l|}{yeast} & 0.065 & 0.100 & 0.144 & 0.221 & 0.006 & 0.022\\  
    \multicolumn{2}{|l|}{agblog} & 0.025 & 0.053 & 0.085 & 0.119 & 0.002 & 0.008\\   
    \multicolumn{2}{|l|}{hep-th} & 1.08 & 1.46 & 1.85 & 2.63 & 0.257 & 0.917\\   
    \multicolumn{2}{|l|}{facebook} & 0.162 & 0.192 & 0.233  & 0.317 & 0.008 & 0.037\\   
    \multicolumn{2}{|l|}{pokec (g)}& 90.9 & 98.1 & 167.3 & 178.7 & 12.5 & 44.9\\   
    \multicolumn{2}{|l|}{pokec (r)}& 93.9 & 108.1 & 167.7 & 186.1 & 36.1 & 139.2\\   
    \hline
  \end{tabular}
\end{table}

\section{Related Work}
Graph-based semi-supervised learning has proven to be an effective method for classification when very few labels are known and has received a lot of attention~\cite{belkin2004semi, joachims2003transductive, talukdar2009new, zhou2004learning, zhu2003semi}.  Most approaches have assumed ``smoothness'' (i.e.~assortativity) of class labels over the graph structure.  This assumption is entirely reasonable considering that these approaches are typically used for independent data for which graphs are artificially constructed based on similarity.  However in network data, where items are related or interacting, links do not always imply similarity. The idea that graph edges could represent either similarity or dissimilarity of node labels was explored in \cite{goldberg2007dissimilarity} and \cite{tong2007semi}, but in both cases it was known a-priori for each edge whether or not it implied similarity of labels.  

Linearised Belief Propagation~\cite{gatterbauer2015linearized}, used as a baseline model in this work, allows for link-heterogeneity by using a label affinity matrix to describe how different classes interact with each other. Similarly, a label propagation approach has been proposed with the added benefit of accurately predicting the confidence of the classification output~\cite{yamaguchi2016camlp}. These methods do not require prior knowledge of which  edges are heterogeneous, but do require the affinity matrix to be specified.  Recent work~\cite{gatterbauer2014Semi} has developed methods for estimating the class affinities from the network. However, as we have shown here, our methods usually perform better even when complete knowledge of the true affinities is provided.  

To the best of our knowledge the only works to account for class-heterogeneity are \cite{peel2011topological, peel2012supervised} which do so by using a stochastic blockmodel (SBM)~\cite{holland1983stochastic, nowicki2001estimation} to model the network structure and learn the relationship between SBM groups and the class labels.  With more SBM groups than class labels, it is possible to capture class-heterogeneity.  However, the computational expense of these methods makes them unsuitable for large networks. 
Furthermore, we demonstrated on synthetic networks that they are less accurate than the new methods  we present here.

\section{Conclusion}
We have considered the problem of node classification in relational networks.  We introduced two novel approaches to graph-based semi-supervised learning based on label propagation, to allow for link-heterogeneity and class-heterogeneity without prior knowledge of how nodes with given class labels interact.  Both methods performed well, but our \textit{two-step label propagation} algorithm gave the best overall performance with respect to accuracy, precision and computational efficiency.

In this work we used node equivalence relations from social network theory and cosine similarity to identify similar graph vertices.  An interesting direction for future research would be to investigate other vertex similarity measures~\cite{Blondel2004Measure, Leicht2006Vertex} and identify, if any, the types of networks they are best suited for.  In~\cite{Blondel2004Measure} they consider a vertex similarity measure for use across different networks, which opens up the possibility of applying some of the ideas developed here to the task of transfer learning in networks~\cite{niuanalyzing}.

\section*{Acknowledgments}
The author would like to thank Jean-Charles Delvenne, Wolfgang Gatterbauer, Cris Moore and Michael Schaub for helpful conversations.  The author was supported by IAP ``DYSCO'' of the Belgian Scientific Policy Office, and ARC ``Mining and Optimization of Big Data Models'' of the Federation Wallonia-Brussels.

%

\bibliographystyle{abbrv}
\bibliography{refs}
%
%
\appendix


\end{document}